# Accurate Evaluation of Asset Pricing Under Uncertainty and Ambiguity of Information

Farouq Abdulaziz Masoudy

College of Business Administration, Taibah University, Saudi Arabia

Email: fmasoudy@taibahu.edu.au

## Abstract

Since exchange economy considerably varies in the market's assets, asset prices have become an attractive research area for investigating and modeling ambiguous and uncertain information in today markets. This paper proposes a new generative uncertainty mechanism based on the Bayesian Inference and Correntropy (BIC) technique for accurately evaluating asset pricing in markets. This technique examines the potential processes of risk, ambiguity, and variations of market information in a controllable manner. We apply the new BIC technique to a consumption asset-pricing model in which the consumption variations are modeled using the Bayesian network model with observing the dynamics of asset pricing phenomena in the data. These dynamics include the procyclical deviations of price, the countercyclical deviations of equity premia and equity volatility, the leverage impact and the mean reversion of excess returns. The key findings reveal that the precise modeling of asset information can estimate price changes in the market effectively.

## 1. Introduction

With the boom of technological advances in our era, the world economy has significantly been increasing, and expectations between multiple investors about future increase are also higher than current growth. It is clearly noted that after the world crisis of 2008, the variations of asset prices have become an important matter in order to estimate and predict the market needs [1][2]. This case leads to raising some questions of how can we evaluate the economic increase, and what are the impacts on consumption and asset pricing. Answering these questions require investigating the variations of asset prices and their dynamics in different merchandises in the world markets.

The perspectives of psychological and financial specialists assert that people judgments and decisions have a great effect in assessing the variations of goods' prices based on the human needs and demands [1][3]. In [4], the authors proved that individual observations and behaviors can properly represent population distributions. This aspect was named "law of small numbers" that can reveal that person's overweight a few numbers of observations. In an asset situation, this could infer that while investors see a company understanding high incomes increase, for instance, they might categorize it as a growing company and reduction incompetently for the regression aspect [5].

There are various economical views [1] [2] [3] [6] [7] for estimating and predicting asset prices in markets. Welch and Goyal [6] stated that the well-known empirical techniques for estimating and predicting asset prices and returns do not outnumber the independent and identical distribution models in the sample and out-of-sample models, and therefore this is not beneficial for investment mechanisms. Cochrane [7] mentioned that reduced out-of-sample routine is not a test in contradiction of the expectedness of asset returns and prices. When several estimation techniques deliver great differences in predictable returns and prices, the extrapolative relationship is unsteady.

There are also many experimental studies in the field that show the trading of persons and investors for revealing the extrapolation of past performance, and how it can affect capital market perspective. Fuster et al. [3] proposed that extrapolation is significant for studying macroeconomic variations. Barberis et al. [8] suggested that over-extrapolation could clarify the 2008 credit crisis. Further to these studies, production-based asset pricing techniques face a big problem than endowment-based techniques in describing consumption and asset return and price aspects, for example, techniques permit greater scope for endogenous consumption and dividend smoothing [1] [9].

Many research studies [17] [14] [13] [15] [16] investigated the impacts of extrapolative expectations. La Porta et al. [17] offered evidence of overreactions, which proposed that extrapolation would assist in describing stylized facts about the predictability of aggregate market prices and returns. Bansal and Shaliastovich [14] examined the impacts of recentness bias in an exchange economy. The authors revealed that extrapolative bias could assist in demonstrating a high equity premium and high stock market volatility.

An investment volatility technique was suggested by Lansing [13], which adjusted capital costs of asset pricing. Because of the capital adjustment costs, the investment increase shows the same volatility as the technique outcome, while the investment increase in the data is approximately two times greater than the technique outcome. Barberis et al. [15] proposed a consumption-based and asset pricing technique, and their outcomes revealed that the stock price extrapolative bias extracts several attributes of real and prices. Collin-Dufresne et al. [16] discussed that the price extrapolative bias could be a major aspect of the asset prices dynamics.

This study differs from the previous as we investigate the impacts of extrapolation asset prices and to what extent ambiguous information could affect it. The Bayesian Inference and Correntropy (BIC) technique has inferences for both price and quantity attributes for accurately evaluating asset pricing in markets. As previously discussed, it is difficult to introduce a huge equity premium when consumption would be smoothed by companies' investment decisions as in the proposed technique. The proposed technique shows that extrapolation can assist in producing more volatility for both the wealth-consumption ratio and aggregate wealth return due to the large deviations in expectations of the technological increase in our era.

The key contributions in this study are declared as follows. Firstly, we suggest a novel uncertainty technique based on the Bayesian Inference and Correntropy theories for precisely estimating asset pricing in markets. Secondly, the proposed technique inspects the inherit phases of uncertainty and variations of market information in a manageable fashion. Thirdly, the technique is utilized for modelling the consumption asset pricing technique in which the consumption deviations are developed by the Bayesian inference with observing the dynamics of asset pricing phenomena in the data, involving the procyclical variations of price, the countercyclical variations of equity premia and equity volatility, the advantage impact and the mean reversion of excess returns.

The rest of the paper is organized as follows. Section 2 discusses the ambiguous and uncertain information of asset pricing. The proposed BIC technique for measuring asset prices and returns are described in section 3. Section 4 describes the data and model estimation. In section 5, the results and discussion related to the proposed BIC technique are provided. Finally, the paper conclusion is explained in Section 6.

## 2. Ambiguous and uncertain information of Asset Pricing

In the financial market, ambiguity is known as the information obtained from unreliable sources in which it could be difficult to judge the quality of critical information [18]. Processing such information without complete and secure/reliable knowledge could lead investors to not go through the economic process and treat this information as ambiguous, and as a result, they cannot evaluate the excess return or any asset outcomes in the same way as with the trusted and reliable information [19].

Ambiguous information mainly influences on the market agents and investors by two main effects in which obtaining such this information lead agents to take asymmetric actions such as bad news affects more than good news in decision making [17]. In addition, ambiguous information could imply agents to expect and aversion the relevant consumption strategies. This effect intuitively could not appear using Bayesian approaches whereas they prevent the dependence of coming information quality and current services and actions [1] [20].

In the literature, others as in [10] present the ambiguity as for the uncertainty of probabilities in which the relevant or associated probabilities to specific situations are not clear or repeated for several situations. In other words, the uncertain probabilities make up risk, which sometimes is identified by ambiguity so that the probabilities associated with some recognized insights should be defined and uniquely assigned for minimizing their risks.

Leippold [28] found that learning upon ambiguous information could provide balance deduction and the interest rates could be affected when the risk aversion is so small. When the risk aversion is too low, learning and ambiguity aversion raise conditional equity premia and volatilities and this equity premium, which is correspondence to the interaction of learning and ambiguity aversion is the dominant part.

Therefore, interpreting market information or signals is so important in order to judge the information quality, seeking to consistent and reliable decisions. This is completely true, especially for what is called "tangible" information, which known as significant and complete information signals that assist in market quantitative analysis such as earning and interest rates/reports [10]. Moreover, the relationship between the excess returns and conditional return variances is greatly time-varying [28].

## 3. Proposed Bayesian Inference and Correntropy (BIC) technique for assessing asset pricing

This section discusses the theories of Bayesian Inference and Correntropy for developing the BIC technique for estimating and predicting the asset pricing under uncertainty information of markets.

### 3.1 Bayesian inference

Bayesian inference [11] is defined as the capability of assigning/allocating probability distributions to any event either the involved process is random or not. In Bayesian approach, probability refers to a quantitative and subjective measure of a person's degree of relief. There is a big difference between Bayesian inference that causes much confusion while understanding the Bayes theory and Bayesian inference. The main difference between them is their goals as the former focuses on quantifying and analyzing believes' degrees while the latter depends on frequency procedures. Therefore, the best choice for use is based on the desired goal.

In Bayes' theory, the conditional probability of an event is obtained with respect to another probability event that has already happened [22]. It expresses the way of estimating the conditional probability or posterior probability of an event *A* given additional information of another event *B* with their prior probabilities and the conditional probabilities of B given A, as shown in the following formula:

$$P(A|B) = \frac{P(B|A).P(A)}{P(B)} \quad (1)$$

Bayesian inference is consequently derived from Bayes' theory by exchanging B with observations *Y, A* with parameter set *Θ* and probabilities *P* with function *f* whereas P(Θ) represents the prior probability distribution of the parameter set Θ before perceiving Y observations, a p(Y|Θ) is the probability of Y based on a mathematical model, and finally

p(Θ|Y) is the posterior probability to be estimated after observing Y, represented in the formula:

$$P(\Theta|Y) = \frac{P(Y|\Theta).P(\Theta)}{P(Y)} \quad (2)$$

## 3.2 Correntropy theory

Correntropy technique [12] is the measure of estimating the degree of similarities between feature vectors/records as it can precisely identify the differences in the behavior of observations. More specifically, it can provide statistical evidence of the significant variations of these features if there is an obvious and distinct difference between these feature records [21]. It is known as one of the second-order, non-linear statistical technique that shows the associations and interactions between these observations. Mathematically speaking, supposing two random variables r1 and r2, their correntropy for showing their relationship is given by

$$V_\sigma(r1, r2) = E[K_\sigma(r1 - r2)] \quad (3)$$

where E[.] denotes the mathematical expectation of feature vectors, $\kappa_\sigma(.)$ refers to the Gaussian kernel function and σ defines the kernel size, presented by

$$K_\sigma = \frac{1}{\sqrt{2\pi\sigma}} \exp\left(-\frac{(.)^2}{2\sigma^2}\right) \quad (4)$$

Since the joint probability density function between these two variables $r_1$ and $r_2$ is usually ambiguous, the number of vectors is known ($\{a_i, b_i\}_{i=1}^{M}$). Subsequently, the formula of computing correntropy in equation (2) is replaced by

$$V_{M,\sigma}(r1, r2) = \frac{1}{M}\sum_{i=1}^{M} K_\sigma(r1_i, r2_i) \quad (5)$$

In order to describe the configuration of applying the BIC technique, the technique is adjusted with extrapolative expectations for determining the dynamics of consumption, investment, output, and asset prices. In more details, the Bayesian inference in subsection 3.1 is applied for estimating the dynamics of pricing changes over time while its output is used as input of the correntropy mechanism in order to exactly reveal the similarity of pricing changes in each time series. When the correntropy outcomes of each two consecutive time series are similar, this means that there is no extrapolative bias between the two.

## 4. Data and model estimation

Since there are a lot of research studies in the literature reporting the predictability of stock returns [23, 24], some of the predictive features/variables are used such as valuation ratios, short rates, the slope of the yield curve, payout ratios and other stock features in the asset prices. Using annual data captured from the U.S. stock marketplace from the year 1927 until the year 2010 for evaluating the variations of market consumption information and checking the asset prices dynamics over the current data.

The log returns and 90 Day T-Bill return sequences of CRSP market portfolio as a way of estimating the annual risk-free proportions. The Consumer Price Index (CPI) diminishes the nominal values, results in the mean risk-free proportion that equals 0.0078. The predictive features are chosen according to the portfolio choice in the literature [23-26], the log difference among the cumulative and excess stock returns of the CRSP yielding the demeaned sequences as shown in Figure 1.

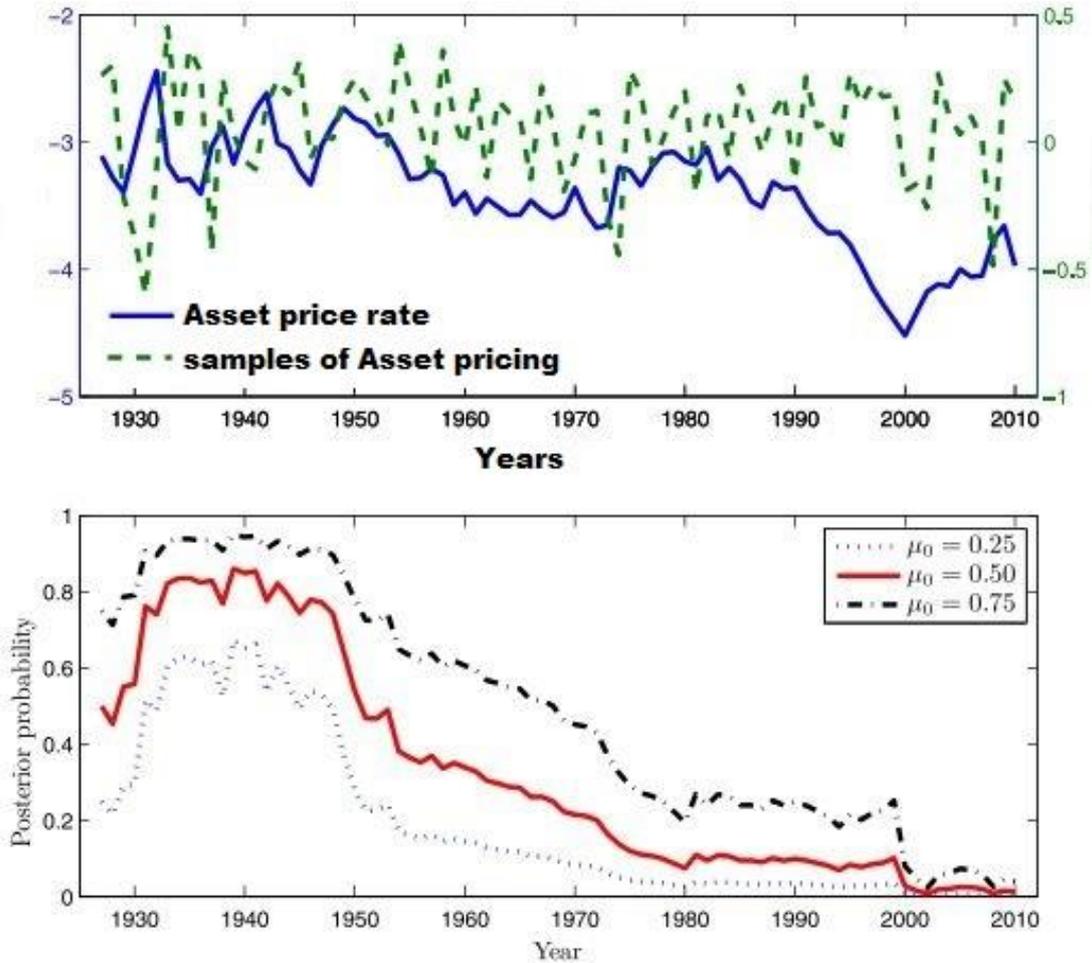

Figure 1: (A) asset price rate and their samples and (B) plots the posterior probabilities of the BIC technique

The second plot in Figure 1 shows the posterior probabilities of the BIC model using the historical stock returns data over the period 1927 till 2010 with four values of the prior probabilities = 0, 0.25, 0.5 and 0.75. From this plot, it can be observed that the overall trend is downward over years reflecting the proposed mechanism can estimate the variations and dynamics of the asset prices and their parameters efficiently.

## 5. Results and discussion

To choose a Bayesian network for modeling data of asset prices, we generate data that build a normal distribution with its parameters: mean and standard deviation to maximize the likelihood of the Byes theorem, as shown in Figure 2. The results in the figure show a potential solution for the two parameters in order to represent the path of posteriors that is totally inspected by the path of price dynamics.

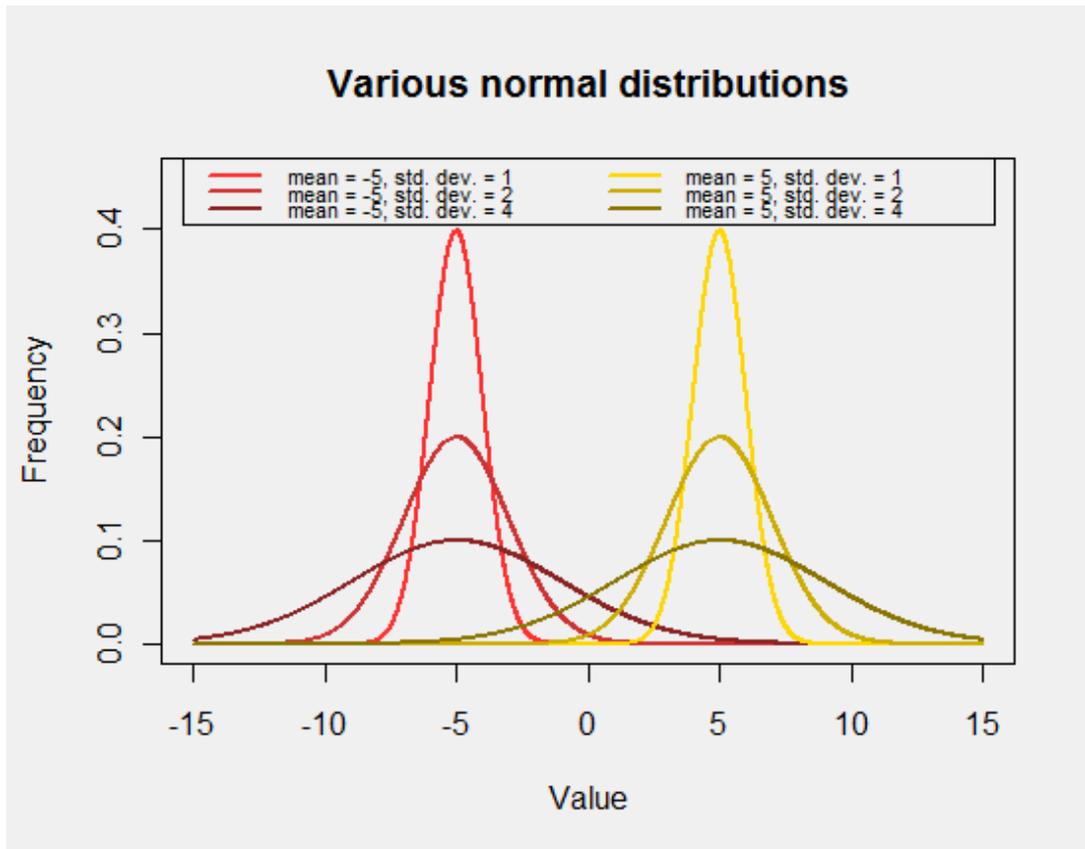

(A)

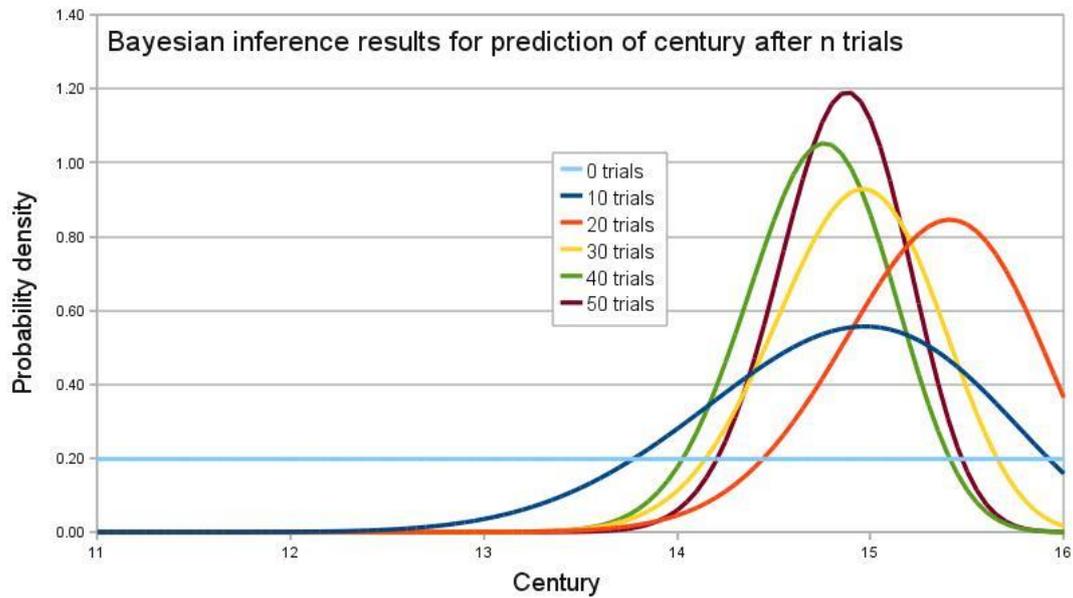

(B)

Figure 2: (A) gives an example of generating a normal distribution of asset pricing data, and (B) presents an example of applying Bayesian inference for this data

The performance differences through Bayesian inferences rely on how possibly the posteriors' path is the truth. If the variation in a term of the square of the standard deviation is very large, then there could have been enough not got news to describe the initial price reduction. On the contrary, if the variation is so small, then signals are irrelevant that posteriors do not considerably change to any given news. This reveals that the asset price dynamics can be managed in terms of good or not good news based on the posteriors of the Bayesian inference that estimated from the asset pricing information with extrapolative expectations, showing the price paths over the time.

By applying the correntropy for the posteriors of the Bayesian networks, the similarities between these variations for each time series can show is the extrapolative bias between each sequenced series. In other words, when the correntropy value asset pricing two series are near two each other in the plot, it means that the asset prices of them do no change significantly, otherwise there is a great change in the prices that we have to consider why this change happened, as shown in Figure 3.

The empirical results reveal how the asset pricing and its attributes can be effectively modeled using the BIC technique. Based on the Bayesian network mechanism, all prices movement displays changes in beliefs about the future increase. Specifically, the initial decrease in prices arose, as market contributors expected a lasting drop in consumption over the time. The term of bad news augmented the predictable fall over the time. On the other hand, the ambiguity and uncertainty information of the asset prices can be estimated using the posteriors of the Bayesian network. This can be done based on agents to start with prior information about asset prices in the market that nothing has altered over the time.

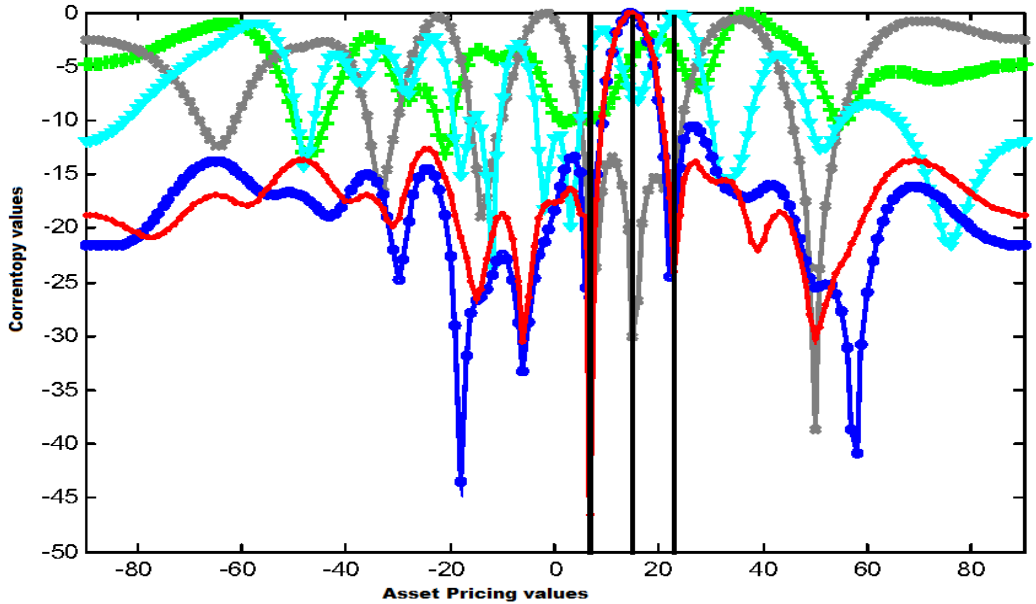

Figure 3: correntropy of different time series data of asset pricing.

Modeling the posteriors using the correntropy can describe the deviations of asset pricing and its features over the time showing the ambiguity of information about what could be occurred in these dynamics in the future. For example, the big difference between two signals of the correntropy demonstrates that there is the uncertainty of asset pricing. For further explanation, the future bad news can be declared the big difference between each two consecutive asset pricing correntropy, while the future good news which can be interpreted by each two consecutive asset pricing correntropy.

## 6. Conclusion

This paper covers a Bayesian Inference and Correntropy (BIC) technique for precisely assessing asset pricing in the daily markets. The technique determines the inherent processes of ambiguous and uncertain information of asset prices. The Bayesian inference approach is used for designing a consumption asset pricing model in order to examine the variations in the data of asset pricing, whereas the correntropy is used for checking the difference that happens between any two time series of asset pricing and creating their features over the time. This reveals that if there will be ambiguous information about the asset pricing's dynamics in the future. The outcomes demonstrate that the utilization of BIC technique can efficiently estimate and predict the changes in asset prices. In future, the proposed technique will be extended to apply for real asset pricing data showing its efficiency.